\documentclass{article}

\usepackage{tikz, pgf, pgfplots}
\usetikzlibrary
	{
		arrows, automata, chains, datavisualization.formats.functions, fit, positioning, quantikz, quotes, shapes
	}

\usepackage{amsthm}
\usepackage{amssymb}
\usepackage[bb = boondox]{mathalfa}
\usepackage{dsfont}
\usepackage{mathtools}
\usepackage{multicol}
\usepackage{braket}
\usepackage{physics}
\numberwithin{equation}{section}

\usepackage{yquant, braket}

\usepackage[a4paper, left = 2.5cm, right = 2.5 cm, top = 2.5cm, bottom = 3.0 cm]{geometry}

\usepackage{xcolor}
\usepackage{varwidth}
\usepackage[breakable, skins, theorems]{tcolorbox}
\usepackage{graphicx}
\usepackage{hyperref}
\hypersetup
{
	colorlinks,
	linkcolor = {WordRed!75!black},
	citecolor = {WordBlueDarker25},
	urlcolor = {WordAquaDarker50}
}
\usepackage{nicematrix}
\NiceMatrixOptions
{
	code-for-first-row = \color{RedPurple},
	code-for-last-col = \color{WordAquaDarker25}
}
\usepackage{colortbl}
\usepackage{float}
\usepackage{cellspace}
\usepackage{makecell}
\usepackage[boxed, ruled, vlined]{algorithm2e}
\usepackage{appendix}
\usepackage{multirow}
\newcommand{\orcidicon}[1]{\href{https://orcid.org/#1}{\includegraphics[height=\fontcharht\font`\B]{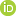}}}
\definecolor{MyLightRed}{RGB}{244, 213, 245}
\definecolor{WordRed}{RGB}{255, 0, 102}
\definecolor{RedDarkLightest}{HTML}{ff0088}
\definecolor{RedDarkLight}{HTML}{ea005f}
\definecolor{RedPurple}{HTML}{aa007f}
\definecolor{Purple}{HTML}{911146}
\definecolor{WordLightGreen}{RGB}{140, 214, 192}
\definecolor{WordGreen}{RGB}{0, 176, 80}
\definecolor{GreenLightest}{HTML}{00ffa0}
\definecolor{GreenLighter1}{HTML}{00b383}
\definecolor{GreenLighter2}{HTML}{00aa7f}
\definecolor{GreenDark}{HTML}{225522}
\definecolor{GreenTeal}{HTML}{008080}
\definecolor{WordIceBlue}{RGB}{223, 227, 229}
\definecolor{MyVeryLightBlue}{RGB}{211, 245, 247}
\definecolor{WordBlueVeryLight}{RGB}{0, 176, 240}
\definecolor{WordBlueLight}{RGB}{0, 112, 192}
\definecolor{WordBlueDark}{RGB}{46, 116, 181}
\definecolor{WordBlueDarker}{RGB}{31, 78, 121}
\definecolor{WordBlueDarker25}{RGB}{54, 96, 146}
\definecolor{WordBlueDarker50}{RGB}{36, 64, 98}
\definecolor{WordBlueDarkest}{RGB}{0, 32, 96}
\definecolor{WordBlue}{RGB}{19, 65, 99}
\definecolor{MyBlue}{RGB}{0, 64, 128}
\definecolor{MyDarkBlue}{RGB}{0, 51, 102}
\definecolor{BlueVeryDark}{HTML}{222255}
\definecolor{WordAquaLighter80}{RGB}{218, 238, 243}
\definecolor{WordAquaLighter60}{RGB}{183, 222, 232}
\definecolor{WordAquaLighter40}{RGB}{146, 205, 220}
\definecolor{WordAquaDarker25}{RGB}{49, 134, 155}
\definecolor{WordAquaDarker50}{RGB}{33, 89, 103}
\definecolor{WordVeryLightTeal}{RGB}{223, 236, 235}
\definecolor{WordLightTeal}{RGB}{160, 199, 197}
\definecolor{WordDarkTealLighter80}{RGB}{207, 223, 234}
\definecolor{WordDarkTeal}{RGB}{72, 123, 119}
\definecolor{WordDarkerTeal}{RGB}{48, 82, 80}
\definecolor{WordTurquoiseLighter80}{RGB}{209, 238, 249}
\definecolor{Brown}{HTML}{666633}

\title
	{
		A symmetric extensible protocol for quantum secret sharing
	}

\author
	{
		Michael Ampatzis$^1$\orcidicon{0000-0002-1570-6722}
		and
		Theodore Andronikos$^1$\orcidicon{0000-0002-3741-1271} \\
		$^1$Department of Informatics, Ionian University, \\
		7 Tsirigoti Square, 49100 Corfu, Greece; \\
		\{p16abat, andronikos\}@ionio.gr \\
	}

\begin{document}

\maketitle

\begin{abstract}
	This paper introduces the Symmetric Extensible Quantum Secret Sharing protocol, which is a novel quantum protocol for secret sharing. At its heart, it is an entanglement based protocol that relies on the use of maximally entangled GHZ tuples, evenly distributed among the players, endowing the spymaster with the ability to securely share a secret message with her agents. It offers uncompromising security, making virtually impossible for a malicious eavesdropper or a rogue double agent to disrupt its successful execution. It is characterized by symmetry, since all agents are treated indiscriminately, utilizing identical quantum circuits. Furthermore, it can be seamlessly extended to an arbitrary number of agents. After the completion of the quantum part of the protocol, the spymaster will have to publicly transmit some information, in order to enable the agents to discover the secret message. Finally, it has the additional advantage that the spymaster has the privilege to decide when it is the right time for the agents to discover the secret message.
	\\
	\\
\textbf{Keywords:}: Quantum secret sharing, quantum cryptography, quantum entanglement, GHZ states.
\end{abstract}
\section{Introduction} \label{sec:Introduction}

After the landmark announcement of IBM's new quantum computer, which managed to successfully break the 100-qubit barrier \cite{IBMEagle}, it appears that we have moved closer than originally anticipated to the realization of practical quantum computation and the definitive transition from the digital era to the quantum era. However, such rapid technological advancements also herald some rather serious issues with our current infrastructure; issues that can pose significant threats to our information security. After the publication of one of the most famous papers in the field of quantum computing by Peter Shor in 1994 \cite{Shor1994}, which proposed an algorithm that can solve the integer factorization problem in polynomial time and is thus theoretically able to break any cryptosystem based on the aforementioned problem, it was made apparent that if quantum computers become a reality, then the existing digital security protocols must be replaced with ones that can withstand attacks from a quantum computer.

One of the numerous sub-fields of cryptography which can be improved with the use of quantum mechanics, in order to enhance its security and efficiency against a quantum computer, is the field of secret sharing, which can be described as a clandestine game between three or more individual players, that are located in two different geographical locations and are unable to communicate in person. For the sake of simplicity, we shall describe the game with only three players, as it is the simplest version. However, as mentioned above, the game can be scaled up to as many players as we want. Therefore, we start by dividing the three players into two groups. The first group consists of the \emph{leader} or the \emph{spymaster}, which is a single person and for this game we shall give this honor to Alice, who is located in Athens. The second group that consists of her \emph{secret agents}, which are the rest of the players, in this case Bob and Charlie who are located in Corfu. In this game the spymaster Alice wants to send a set of secret instructions to her agents for them to act upon. However, Alice is not certain that she can trust both of her agents because she suspects that one of them, but not both, is a double agent and is working for the enemy, with the goal of foiling her plans. Although, she knows that if both of her agents partake in the mission together, the agent that is loyal to Alice will prevent the double agent from causing any damage. Furthermore, aside from the possible double agent who wants to sabotage the mission, Alice must also take precautions to assure the confidentiality of the mission from possible eavesdroppers, which for the sake of the game will be referred to as Eve. Therefore, Alice must somehow find a way to make both her agents act on her behalf, and, at the same time, prevent both the dishonest agent and Eve from tempering with the mission.

This elaborate game may at first sight seem a little redundant or even superfluous and yet, the field of secret sharing, has been proven vital for a certain type of problems with multiple real world uses, as for example, to guarantee that a single individual or a small group of the involved party can not access a valuable shared resource, such as a shared bank account. This means that in order to take any action, all involved members must act in concert. As a consequence, this requirement makes it considerably more difficult for any individual who wants to have unauthorized access to a secret information or wants to perform an unauthorized action, to achieve her purpose. Practically, this implies that the only way for a malicious player to perform any action, is to convince every single member to go along.

The pioneering works of Hillery et al. \cite{hillery1999quantum} and Cleve et al. \cite{cleve1999share}, which proved that secret sharing can be achieved with the use of quantum mechanics, paved the way for the creation of a new field with the name of Quantum Secret Sharing (QSS for short). Since then, there has been extensive progress in this field, with a plethora of proposals actively continuing the research to this day \cite{karlsson1999quantum, smith2000quantum, gottesman2000theory, bandyopadhyay2000teleportation, xiao2004efficient, fortescue2012reducing, qin2020hierarchical, senthoor2022theory}. Furthermore, several experimental demonstrations have been made, by multiple research groups, proving the viability of such protocols, even in a real world scenario \cite{tittel2001experimental, bogdanski2008experimental, bell2014experimental}. Another line of research attempted to extend the capabilities of the field, by proposing non-binary protocols that rely on the use of qudits rather than qubits \cite{tavakoli2015secret, pinnell2020experimental}. With this rapid advancement of the field, one may also observe some notable efforts towards the implementation of Grover's algorithm \cite{Grover1996}, which is one of the most famous algorithms of all time, as a QSS protocol \cite{hsu2003quantum, hao2011quantum, yu2022improved}.

In this paper we propose a new QSS protocol in the form of a quantum game, which we hope will make its presentation simple and pedagogical, thanks to the inherent nature of game theory to present complex and challenging problems in a more methodical and comprehensive way. As a consequence quantum games have gained prominence as a useful paradigm of the quantum world and have been used to tackle important problems. A typical example in this area is the quantum game of coin tossing and its application to cryptographic protocols (see \cite{Bennett1984}, \cite{Bennett2014} and the more recent \cite{Ampatzis2021}). The beginning of this field can be traced back to two landmark papers from 1999: Meyer's PQ penny flip game \cite{Meyer1999} and the Eisert-Wilkens-Lewenstein scheme \cite{Eisert1999}. Numerous works inspired from the PQ penny flip game have appeared in the literature and some recent results are given in \cite{Andronikos2018, Andronikos2021, Andronikos2022}. The Eisert-Wilkens-Lewenstein scheme was successfully employed in the study of quantum versions of well-known classical games, such as the famous Prisoners' Dilemma, where the quantum strategies outperformed the classical strategy \cite{Eisert1999}, and to quantum extensions of the classical repeated Prisoners' Dilemma strategies \cite{Giannakis2019}. Winning strategies for abstract quantum games can also be encoded as infinite words accepted by quantum automata, as demonstrated in \cite{Giannakis2015a}. Quantum games have even be used in \cite{Andronikos2022} as a metaphor for the operation of a hypothetical quantum parliament. Striving to outperform traditional methods when dealing with problems and games, motivated researchers to also turn to the biological domain. Many classical games, including the Prisoners' Dilemma, have also been examined under the prism of biological notions \cite{Kastampolidou2020, Kastampolidou2021, Kastampolidou2021a, Papalitsas2021}. Unsurprisingly, computation can be conducted in the context of biological and bio-inspired processes \cite{Theocharopoulou2019, Kastampolidou2020a}, even including computer viruses \cite{Kostadimas2021}.

\textbf{Contribution}. As a result of our approach, in this paper we propose the novel Symmetric Extensible Quantum Secret Sharing protocol, which we designate by SEQSS$_n$, where $n$ represents the number of participating players. The final result is an entanglement based protocol that relies on the use of maximally entangled GHZ tuples, evenly distributed among the players, in order to allow our spymaster to securely share the secret message with her agents.  As the name suggests it exhibits symmetry, since all agents are treated indiscriminately, utilizing identical quantum circuits. It offers uncompromising security, making virtually impossible for a malicious eavesdropper or a rogue double agent to disrupt its successful execution. Additionally, after the completion of the quantum part of the protocol, it is mandatory for the spymaster to publicly communicate a piece of information to anyone of the agents, since without this information the agents will not be able to discover the secret message, even if they combine together all their individual data. However, this can be considered as an added advantage, because the spymaster has the initiative to decide, when it is the right time for the agents to read the message. For the sake of simplicity and enhanced readability, we shall initially describe a simplified version of the game with only three players. One of the main traits of the proposed protocol is its extensibility, meaning that the game of secret sharing can be seamlessly scaled up to as many players as we want. Hence, we shall afterwards present the protocol in its most general form, involving $n$ players in total.  

\subsection{Organization}

This paper is organized as follows. Section \ref{sec:Introduction} gives an introduction to the subject along with some relevant references. Section \ref{sec:Background on GHZ States} provides a brief introduction about the GHZ states and the principle of entanglement, a vital tool used in the formulation of the proposed protocol. Section \ref{sec:The Symmetric Extensible QSS Protocol} showcases the Symmetric Extensible QSS protocol twice, once in its simplest form, which is restricted to only three players and once in its general form, which is played by $n$ players.
Finally, Section \ref{sec:Discussion and Conclusions} summarizes and discusses the functionality and advantages of the proposed protocol.

\section{Background on GHZ states} \label{sec:Background on GHZ States}

Quantum entanglement can be described mathematically as the linear combination of two or more product states. This unique phenomenon is universally considered as one of the fundamental principles of quantum mechanics and is the basis of many fascinating proposals, one of which is the achievement of quantum teleportation. Moreover, entanglement based techniques have been proven indispensable in the field of quantum secret sharing and quantum cryptography in general. In this paper, for the realization of our QSS protocol, we will heavily rely on the utilization of maximally entangled pairs of three or more qubits, which are commonly known as GHZ states. Therefore, we believe that the following subsection, in which we give a brief explanation on the nature of GHZ states, will facilitate the understanding of the proposed protocol.

\begin{figure}[H]
	\begin{tcolorbox}
		[
			grow to left by = 1.75 cm,
			grow to right by = 1.75 cm,
			colback = white,		
			enhanced jigsaw,		
			sharp corners,
			toprule = 1.0 pt,
			bottomrule = 1.0 pt,
			leftrule = 0.1 pt,
			rightrule = 0.1 pt,
			sharp corners,
			center title,
			fonttitle = \bfseries
		]
		\centering
		\begin{minipage}[t]{0.475 \textwidth}
			\includegraphics[scale = 0.4, trim = {0 0 2cm 0}, clip]{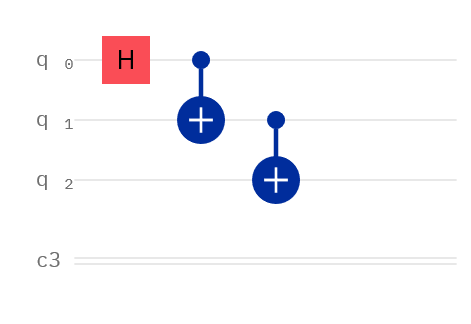}
			\caption{This quantum circuit can be used in Qiskit to entangle 3 qubits in the $\ket{ GHZ_3 } = \frac{ \ket{0} \ket{0} \ket{0} + \ket{1} \ket{1} \ket{1} } {\sqrt{2}}$ state.}
			\label{fig:GHZ_3_QC}
		\end{minipage}
		\hfill
		\begin{minipage}[t]{0.475 \textwidth}
			\includegraphics[scale = 0.3, trim = {0 6cm 16cm 0cm}, clip]{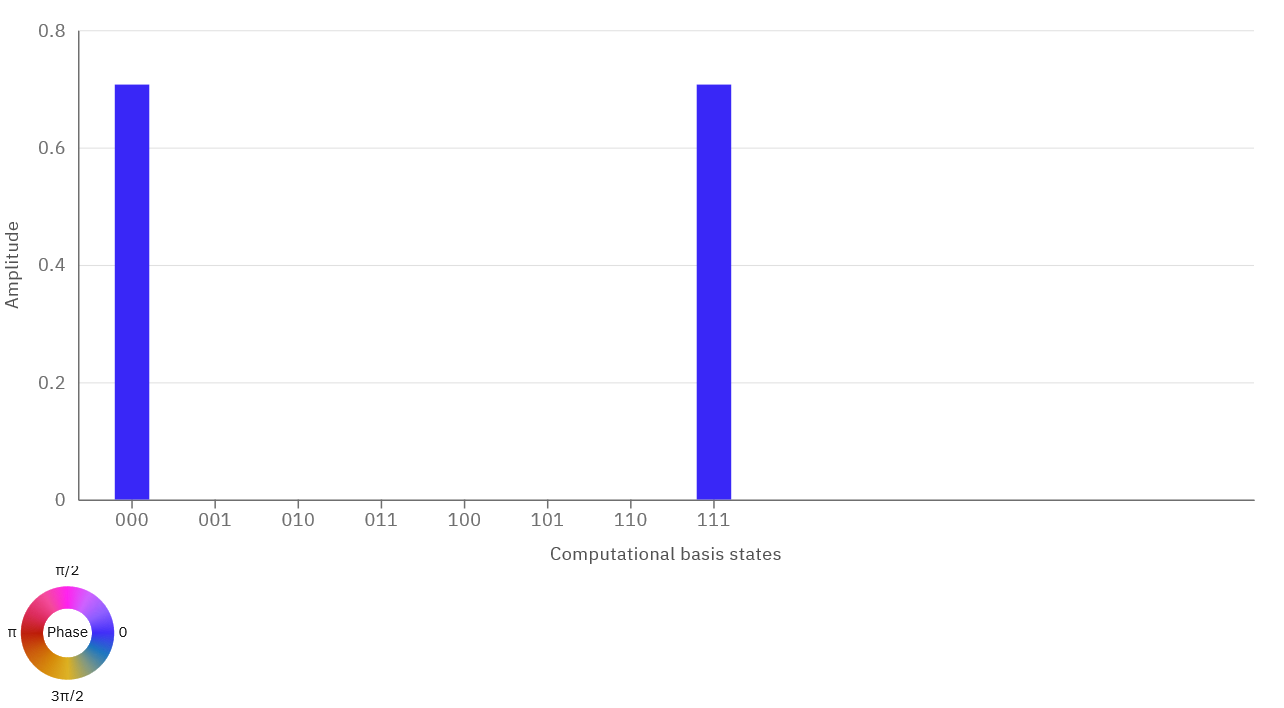}
			\caption{The state vector description of 3 qubits entangled in the $\ket{ GHZ_3 }$ state.}
			\label{fig:GHZ_3_SV}
		\end{minipage}
	\end{tcolorbox}
\end{figure}

From the perspective of quantum computing, a GHZ state can be produced by a circuit with three or more qubits, upon which a Hadamard gate is applied to the first qubit, then a CNOT gate is applied to the first and second qubits, and subsequently a CNOT to the second and third qubits, and so on until we reach the last qubit. A possible quantum circuit for preparing three qubits in the GHZ state is shown in Figure \ref{fig:GHZ_3_QC}. Figure \ref{fig:GHZ_3_SV} gives the state vector description of the corresponding GHZ state. These figures were also obtained from IBM Quantum Composer \cite{IBMQuantumComposer2022}.

Using the same reasoning as in the case of the $3$ qubit GHZ state shown in Figure \ref{fig:GHZ_3_QC}, the mathematical representation of a composite system consisting of $n$ single qubit subsystems all entangled in the GHZ state, denoted by $\ket{ GHZ_{n} }$, goes as follows:

\begin{align} \label{eq:Extended General GHZ_n State}
	\ket{ GHZ_{n} } = \frac{ 1 }{ \sqrt{2} }
	\left( \ket{0}_{ n - 1 } \ket{0}_{ n - 2 } \dots \ket{0}_{ 0 } + \ket{1}_{ n - 1 } \ket{1}_{ n - 2 } \dots \ket{1}_{ 0 } \right)
	\ .
\end{align}

The above setting can be generalized to the case where a composite system consists of $n$ subsystems, say $n$ quantum registers $r_0, r_1 \dots, r_{ n - 1 }$, each of them having $m$ qubits, and the corresponding qubits of all the $n$ registers are entangled in the $\ket{ GHZ_{n} }$ state. The state of the composite system, denoted by $\ket{ GHZ_{n} }^{\otimes m}$, is described by the formula below

\begin{align} \label{eq:m Extended General GHZ_n States}
	\ket{ GHZ_{n} }^{\otimes m}
	&=
	\frac{1}{ \sqrt{2^m} }
	\sum_{\mathbf{x} \in \{ 0, 1 \}^m}
	\ket{\mathbf{x}}_{ n - 1 } \dots \ket{\mathbf{x}}_{ 0 }
	\ .
\end{align}

The proof for this formula is rather straightforward. If we take the entangled state $\ket{GHZ_{n}}$, as shown in (\ref{eq:Extended General GHZ_n State}) twice, the resulting tensor product can be portrayed as

\begin{align} \label{eq: Two GHZ_n states}
	\ket{GHZ_{n}}^{\otimes 2}
	&=
	\frac{1}{ \sqrt{2} }
	\left( \ket{0}_{n-1} \cdots \ket{0}_{0} + \ket{1}_{n-1} \cdots \ket{1}_{0} \right)
	\frac{1}{ \sqrt{2} }
	\left( \ket{0}_{n-1} \cdots \ket{0}_{0} + \ket{1}_{n-1} \cdots \ket{1}_{0} \right)
	\nonumber \\
	&=
	\frac{1}{2}
	\left( \ket{00}_{n-1} \cdots \ket{00}_{0} + \ket{01}_{n-1} \cdots \ket{01}_{0} + \ket{10}_{n-1} \cdots \ket{10}_{0} + \ket{11}_{n-1} \cdots \ket{11}_{0} \right)
	\ .
\end{align}


Following this pattern, we can successfully derive the general formula (\ref{eq:m Extended General GHZ_n States}) with induction and thus complete the proof.

\section{The Symmetric Extensible QSS protocol} \label{sec:The Symmetric Extensible QSS Protocol}

In this section, we thoroughly analyze the proposed Symmetric Extensible QSS (SEQSS for short) protocol in great detail. All of Alice's agents employ \emph{identical} quantum circuits, hence the term symmetric. Due to the nature of the game, the least number of agents Alice can deploy is two, e.g., agents Bob and Charlie, but their number can seamlessly increase to accommodate as many as necessary, which explains the term \emph{extensible}. We begin our analysis with the simplest version, which is a demonstration of the protocol restricted to only three players, designated by SEQSS$_3$, while later on we transition to the general version that consists of $n$ agents and is aptly designated as SEQSS$_n$.

\subsection{The $3$-player SEQSS$_3$ protocol} \label{sec:The 3-Player SEQSS$_3$ Protocol}

Let us begin by showcasing the protocol as a game between three players divided into two groups, exactly as we outlined in Section \ref{sec:Introduction}. Thus, we will have Alice being part of the first group and playing the role of the spymaster, tasked with sharing the secret instructions denoted by $\mathbf{s}$ to her agents, and Bob and Charlie being part of the second group and playing the role of said secret agents, who are spatially separated form Alice, but the are both located in the same region of space. The game will start by having all three players share $m$ triplets of maximally entangled qubits. Each triplet is in the $\ket{ GHZ_{3} }$ state and Alice, Bob and Charlie possess the qubit $\ket{ q_{2} }$, $\ket{ q_{1} }$ and $\ket{ q_{0} }$, respectively, of the triplet. At this point, it is important to state that there are no limitations on which player will create the GHZ triplets in the first place. The states can be created and distributed accordingly by either Alice, her agents, or even by a third party source, i.e., a satellite \cite{aspelmeyer2003long}. The goal of the current game is for Alice to successfully share her secret message $\mathbf{s}$ with her agents Bob and Charlie. However, the secret message $\mathbf{s}$, should not be readable by her agents individually. They must both combine their results in order to be able to read it. This task, can be successfully accomplished, by performing the steps shown below graphically in the form of a quantum circuit.

Furthermore, we point out that each of our three protagonists has a private quantum circuit and the three quantum circuits are similar, in the sense that they all contain an Input Register consisting of $m$ qubits. Bob and Charlie's circuits are virtually identical, whereas Alice's circuit differs because it also contains an Output Register consisting of a single qubit, a Hadamard gate acting on the Output Register and the unitary transform $U_{f_{A}}$, which acts on both her Input and Output Registers. The corresponding qubits in Alice's, Bob's and Charlie's Input Registers constitute a triplet entangled in the $GHZ_3$ state. Table \ref{tbl:Figure SEQSS$_3$ Abbreviationss} explains the abbreviations that are used in the quantum circuit depicted in Figure \ref{fig:The SEQSS$_3$ Protocol}.
-
\begin{tcolorbox}
	[
		grow to left by = 0.75 cm,
		grow to right by = 0.75 cm,
		colback = gray!03,
		enhanced jigsaw,		
		sharp corners,
		toprule = 1.0 pt,
		bottomrule = 1.0 pt,
		leftrule = 0.1 pt,
		rightrule = 0.1 pt,
		sharp corners,
		center title,
		fonttitle = \bfseries
	]
	\centering
	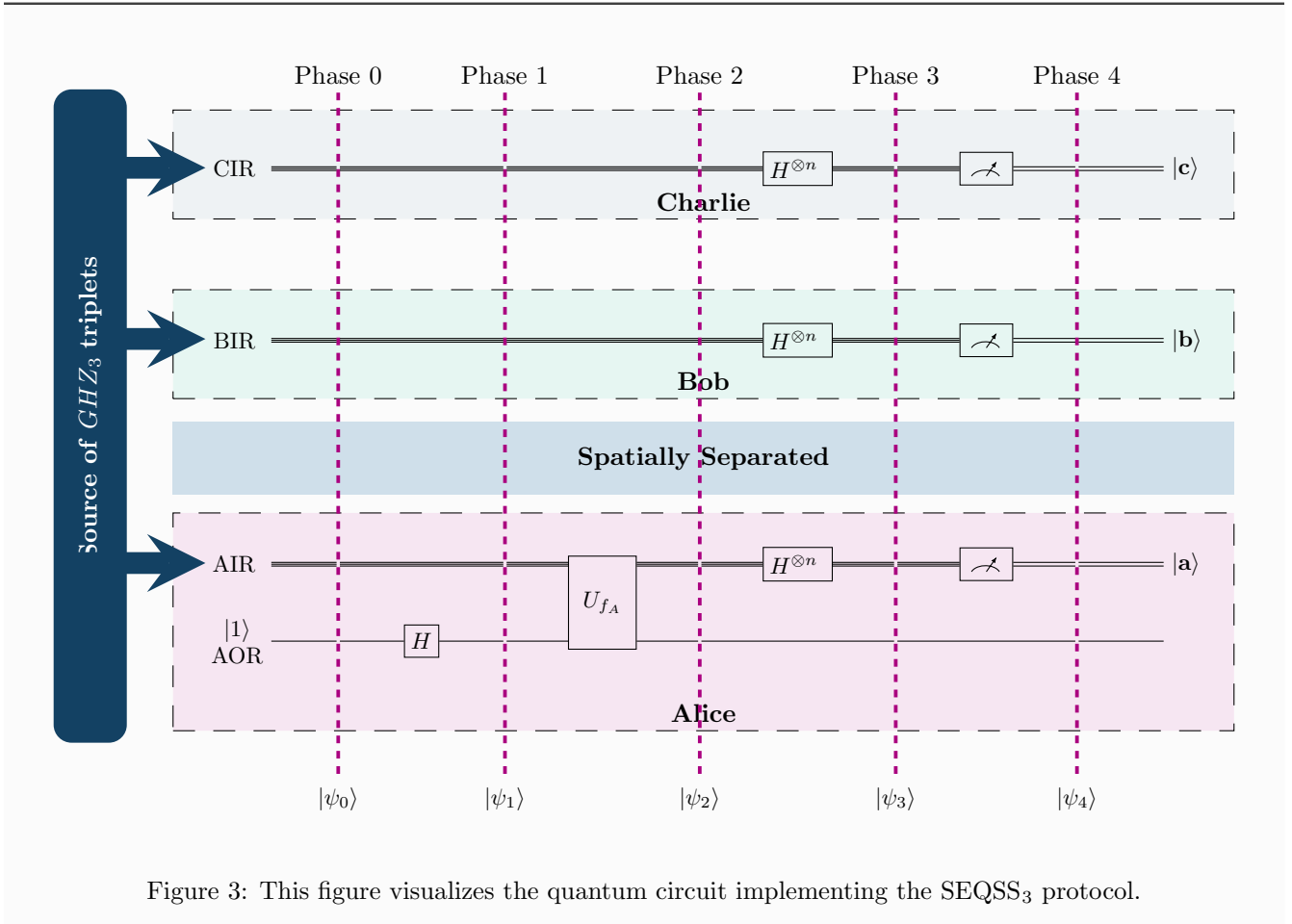
\begin{figure}[H]
		\begin{center}
			\begin{tikzpicture} [ scale = 0.90 ]
				\begin{scope}[on background layer]
					\node
					[
						rectangle, fill = WordBlue!7, draw = black, dash pattern = on 0.30 cm off 0.25 cm, line width = 0.40 pt, minimum width = 145 mm, minimum height = 15 mm, anchor = west, label = { [ anchor = south ] south:\textbf{Charlie} }
					]
					(Charlie) at ( -1.5, -1.0 ) {};
					\node
					[
						rectangle, fill = GreenLighter2!10, draw = black, dash pattern = on 0.30 cm off 0.25 cm, line width = 0.40 pt, minimum width = 145 mm, minimum height = 15 mm, anchor = west, label = { [ anchor = south ] south:\textbf{Bob} }
					]
					(Bob) at ( -1.5, -3.75 ) {};
					\node
					[
						rectangle, fill = WordDarkTealLighter80, minimum width = 145 mm, minimum height = 10 mm, anchor = west, label = { [ anchor = center ] center:\textbf{Spatially Separated} }
					]
					(Space) at ( -1.5, -5.50 ) {};
					\node
					[
						rectangle, fill = RedPurple!10, draw = black, dash pattern = on 0.30 cm off 0.25 cm, line width = 0.40 pt, minimum width = 145 mm, minimum height = 30 mm, anchor = west, label = { [ anchor = south ] south:\textbf{Alice} }
					]
					(Alice) at ( -1.5, -8.00 ) {};
					\node
					[
						rectangle,  rotate = 90, rounded corners = 7 pt, fill = WordBlue, text = white, minimum width = 90 mm, minimum height = 10 mm
					]
					(Source) at (-2.75, -4.85) { \quad \textbf{Source of} $GHZ_3$ \textbf{triplets} \quad };
					\draw [WordBlue, ->, >=stealth, line width = 3.0 mm] (-3.00, -1.05) -- (-1.00, -1.05);
					\draw [WordBlue, ->, >=stealth, line width = 3.0 mm] (-2.60, -3.67) -- (-1.00, -3.67);
					\draw [WordBlue, ->, >=stealth, line width = 3.0 mm] (-3.00, -7.10) -- (-1.00, -7.10);
				\end{scope}
				\begin{yquant}[ operator/separation = 0.30 cm, register/separation = 0.50 cm, every label/.append style = {align = center} ]
					%
					%
					nobit CAUX0;
					qubits { CIR } CIR;
					nobit CAUX1;
					%
					%
					nobit BAUX0;
					qubits { BIR } BIR;
					nobit BAUX1;
					%
					%
					nobit AAUX0;
					nobit AAUX1;
					qubits { AIR } AIR;
					qubit { $\ket{1}$ \\ AOR } AOR;
					nobit AAUX3;
					nobit AAUX4;
					[ name = Ph0, RedPurple, line width = 0.50 mm, label = Phase 0 ]
					barrier ( - ) ;
					h AOR; 
					[ name = Ph1, RedPurple, line width = 0.50 mm, label = Phase 1 ]
					barrier ( - ) ;
					%
					box {$ \ U_{f_{A}} \ $} (AIR, AOR);
					[ name = Ph2, RedPurple, line width = 0.50 mm, label = Phase 2 ]
					barrier ( - ) ;
					box { $H^{\otimes n}$ } CIR, BIR, AIR;
					[ name = Ph3, RedPurple, line width = 0.50 mm, label = Phase 3 ]
					barrier ( - ) ;
					measure CIR, BIR, AIR;
					[ name = Ph4, RedPurple, line width = 0.50 mm, label = Phase 4 ]
					barrier ( - ) ;
					hspace {0.3 cm} CIR;
					output {$\ket{ \mathbf{c} }$} CIR;
					output {$\ket{ \mathbf{b} }$} BIR;
					output {$\ket{ \mathbf{a} }$} AIR;
					\node [ below = 5.25 cm ] at (Ph0) {$\ket{\psi_{0}}$};
					\node [ below = 5.25 cm ] at (Ph1) {$\ket{\psi_{1}}$};
					\node [ below = 5.25 cm ] at (Ph2) {$\ket{\psi_{2}}$};
					\node [ below = 5.25 cm ] at (Ph3) {$\ket{\psi_{3}}$};
					\node [ below = 5.25 cm ] at (Ph4) {$\ket{\psi_{4}}$};
				\end{yquant}
			\end{tikzpicture}
		\end{center}
		\caption{This figure visualizes the quantum circuit implementing the SEQSS$_3$ protocol.}
		\label{fig:The SEQSS$_3$ Protocol}
	\end{figure}
\end{tcolorbox}

\begin{table}[H]
	\renewcommand{\arraystretch}{1.40}
	\begin{tcolorbox}
		[
			grow to left by = -1.00 cm,
			grow to right by = -1.00 cm,
			colback = gray!03,
			enhanced jigsaw, 
			sharp corners,
			boxrule = 0.1 pt,
			toprule = 0.1 pt,
			bottomrule = 0.1 pt
		]
		\caption{This table shows the abbreviations that are used in Figure \ref{fig:The SEQSS$_3$ Protocol}. \\}
		\label{tbl:Figure SEQSS$_3$ Abbreviationss}
		\centering
		\begin{tabular}
			{
				>{\centering\arraybackslash} m{3.00 cm} !{\vrule width 1.25 pt}
				>{\centering\arraybackslash} m{5.00 cm} !{\vrule width 0.5 pt}
				>{\centering\arraybackslash} m{3.00 cm}
			}
			\Xhline{4\arrayrulewidth}
			\multicolumn{3}{c}{\textbf{Abbreviations} used in Figure \ref{fig:The SEQSS$_3$ Protocol}}
			\\
			\Xhline{\arrayrulewidth}
			\textbf{Abbreviations}
			&
			Full name
			&
			\# of qubits
			\\
			\Xhline{3\arrayrulewidth}
			AIR
			&
			Alice's Input Register
			&
			$m$
			\\
			\Xhline{1\arrayrulewidth}
			AOR
			&
			Alice's Output Register
			&
			$1$
			\\
			\Xhline{1\arrayrulewidth}
			BIR
			&
			Bob's Input Register
			&
			$m$
			\\
			\Xhline{1\arrayrulewidth}
			CIR
			&
			Charlie's Input Register
			&
			$m$
			\\
			\Xhline{4\arrayrulewidth}
		\end{tabular}
	\end{tcolorbox}
	\renewcommand{\arraystretch}{1.0}
\end{table}

To describe the resulting composite system we invoke formula (\ref{eq:m Extended General GHZ_n States}) setting $n = 3$. Following the steps of the circuit shown above, we can examine the phases of the algorithm more closely, starting with the initial state $\ket{\psi_0}$ of the system:

\begin{align} \label{eq:SEQSS$_3$ Phase 0}
	\ket{ \psi_0 }
	=
	\frac{1}{ \sqrt{2^m} }
	\sum_{ \mathbf{x} \in \{ 0, 1 \}^m }
	\ket{1}_{A} \ket{\mathbf{x}}_{A} \ket{\mathbf{x}}_{B} \ket{\mathbf{x}}_{C}
	\ .
\end{align}

As expected, $\ket{\mathbf{x}}_{A}$, $\ket{\mathbf{x}}_{B}$ and $\ket{\mathbf{x}}_{C}$ correspond to Alice's, Bob's and Charlie's Input Registers, respectively, representing each player's part of the GHZ states. Likewise, $\ket{1}_A$ represents Alice's Output Register. During our analysis the subscripts $A$, $B$ and $C$ will be consistently used to designate the registers belonging to Alice, Bob and Charlie, respectively. We also emphasize that for consistency we use the Qiskit \cite{Qiskit2022} convention of ordering qubits, where the most significant qubit is the bottom qubit and the least significant qubit is the top qubit.

Continuing to the next phase, Alice effectively initiates the protocol by applying the Hadamard transform to her Output Register. In view of the fact that $H \ket{1} = \frac{1}{\sqrt{2}} ( \ket{0} - \ket{1} ) = \ket{-}$, this produces the ensuing state

\begin{align} \label{eq:SEQSS$_3$ Phase 1}
	\ket{\psi_1}
	=
	\frac{1}{ \sqrt{2^m} }
	\sum_{ \mathbf{x} \in \{ 0, 1 \}^m }
	\ket{-}_{A} \ket{\mathbf{x}}_{A} \ket{\mathbf{x}}_{B} \ket{\mathbf{x}}_{C}
	\ .
\end{align}

This will allow Alice to apply her function $f_{A}$ on her registers by using the standard scheme

\begin{align} \label{eq:SEQSS$_3$ Oracle}
	U_{ f_{A} } : \ket{ y, \mathbf{x} } \rightarrow \ket{ y \oplus f_{A} ( \mathbf{x} ), \mathbf{x} } \ .
\end{align}

Consequently, the next state becomes

\begin{align} \label{eq:SEQSS$_3$ Phase 2 - I}
	\ket{\psi_2}
	=
	\frac{1}{ \sqrt{2^m} }
	\sum_{\mathbf{x} \in \{ 0, 1 \}^m}
	( - 1 )^{ f_{A} ( \mathbf{x} ) } \ket{-}_{A} \ket{\mathbf{x}}_{A} \ket{\mathbf{x}}_{B} \ket{\mathbf{x}}_{C} \ .
\end{align}

At this point let us recall that Alice's function is

\begin{align} \label{eq:Alice's Function}
	f_{ A } ( \mathbf{x} ) = \mathbf{s} \cdot \mathbf{x} \bmod 2 \ ,
\end{align}

where $\mathbf{s}$ is the secret message chosen by Alice. Based on (\ref{eq:Alice's Function}), (\ref{eq:SEQSS$_3$ Phase 2 - I}) can be written as

\begin{align} \label{eq:SEQSS$_3$ Phase 2 - II}
	\ket{\psi_2}
	&=
	\frac{ 1 }{ \sqrt{ 2^m } }
	\sum_{ \mathbf{x} \in \{ 0, 1 \}^m }
	( - 1 )^{ \mathbf{s} \cdot \mathbf{x} }
	\ket{-}_{A}
	\ket{\mathbf{x}}_{A} \ket{\mathbf{x}}_{B} \ket{\mathbf{x}}_{C}
	\ .
\end{align}

Subsequently, Alice, Bob and Charlie apply the $m$-fold Hadamard transform to their Input Registers, driving the system into the next state

\begin{tcolorbox}
	[
		grow to left by = 2.00 cm,
		grow to right by = 1.00 cm,
		colback = white, 
		enhanced jigsaw, 
		sharp corners,
		boxrule = 0.01 pt,
		toprule = 0.01 pt,
		bottomrule = 0.01 pt
	]
	{\small
	\begin{align} \label{eq:SEQSS$_3$ Phase 3 - I}
		\ket{\psi_3}
		&=
		\frac{ 1 }{ \sqrt{ 2^m } }
		\sum_{ \mathbf{x} \in \{ 0, 1 \}^m }
		( - 1 )^{ \mathbf{s} \cdot \mathbf{x} }
		\ket{-}_{A}
		H^{ \otimes m } \ket{ \mathbf{x} }_{A}
		H^{ \otimes m } \ket{ \mathbf{x} }_{B}
		H^{ \otimes m } \ket{ \mathbf{x} }_{C}
		\nonumber \\
		&=
		\frac{ 1 }{ \sqrt{ 2^m } }
		\sum_{ \mathbf{x} \in \{ 0, 1 \}^m }
		( - 1 )^{ \mathbf{s} \cdot \mathbf{x} }
		\ket{-}_{A}
		\left( \frac{ 1 }{ \sqrt{ 2^m } } \sum_{ \mathbf{a} \in \{ 0, 1 \}^m } ( - 1 )^{ \mathbf{a} \cdot \mathbf{x} } \ket{ \mathbf{a} }_{A} \right)
		\left( \frac{ 1 }{ \sqrt{ 2^m } } \sum_{ \mathbf{b} \in \{ 0, 1 \}^m } ( - 1 )^{ \mathbf{b} \cdot \mathbf{x} } \ket{ \mathbf{b} }_{B} \right)
		\left( \frac{ 1 }{ \sqrt{ 2^m } } \sum_{ \mathbf{c} \in \{ 0, 1 \}^m } ( - 1 )^{ \mathbf{c} \cdot \mathbf{x} } \ket{ \mathbf{c} }_{C} \right)
		\nonumber \\
		&=
		\frac{ 1 }{ ( \sqrt{ 2^m } )^{4} }
		\sum_{ \mathbf{x} \in \{ 0, 1 \}^m }
		\sum_{ \mathbf{a} \in \{ 0, 1 \}^m }
		\sum_{ \mathbf{b} \in \{ 0, 1 \}^m }
		\sum_{ \mathbf{c} \in \{ 0, 1 \}^m }
		( - 1 )^{ ( \mathbf{s} \oplus \mathbf{a} \oplus \mathbf{b} \oplus \mathbf{c} ) \cdot \mathbf{x} }
		\ket{-}_{A}
		\ket{ \mathbf{a} }_{A} \ket{ \mathbf{b} }_{B} \ket{ \mathbf{c} }_{C}
		\ .
	\end{align}
	}
\end{tcolorbox}

We can observe now that when

\begin{align}
	\mathbf{a} \oplus \mathbf{b} \oplus \mathbf{c} = \mathbf{s} \ ,
\end{align}

then $\forall  \mathbf{x} \in \{ 0, 1 \}^m$, the expression $( - 1 )^{ ( \mathbf{s} \oplus \mathbf{a} \oplus \mathbf{b} \oplus \mathbf{c} ) \cdot \mathbf{x} }$ becomes $( - 1 )^{0} = 1$. As a result, the sum $\sum_{ \mathbf{x} \in \{ 0, 1 \}^m } ( - 1 )^{ ( \mathbf{s} \oplus \mathbf{a} \oplus \mathbf{b} \oplus \mathbf{c} ) \cdot \mathbf{x} }$ $=$ $2^{m}$. Whenever $\mathbf{a} \oplus \mathbf{b} \oplus \mathbf{c} \neq \mathbf{s}$, the sum is just $0$ because for exactly half of the inputs $\mathbf{x}$ the exponent will be $0$ and for the remaining half the exponent will be $1$. Hence, one may write that

\begin{align} \label{eq:Inner Product Modulo 2 Property}
	\sum_{ \mathbf{x} \in \{ 0, 1 \}^m }^{}
	( - 1 )^{ ( \mathbf{s} \oplus \mathbf{a} \oplus \mathbf{b} \oplus \mathbf{c} ) \cdot \mathbf{x} }
	=
	2^{m} \delta_{ \mathbf{s}, \mathbf{a} \oplus \mathbf{b} \oplus \mathbf{c} }
	\ .
\end{align}

Utilizing equation (\ref{eq:Inner Product Modulo 2 Property}), and ignoring for a moment Alice's Output Register, which is at state $\ket{-}_{A}$, the following three equivalent and symmetric forms can be derived.

\begin{tcolorbox}
	[
		grow to left by = 1.75 cm,
		grow to right by = 0.50 cm,
		colback = white, 
		enhanced jigsaw, 
		sharp corners,
		boxrule = 0.01 pt,
		toprule = 0.01 pt,
		bottomrule = 0.01 pt
	]
	{\small
	\begin{align}
		\sum_{ \mathbf{a} \in \{ 0, 1 \}^m }
		\sum_{ \mathbf{b} \in \{ 0, 1 \}^m }
		\sum_{ \mathbf{c} \in \{ 0, 1 \}^m }
		\sum_{ \mathbf{x} \in \{ 0, 1 \}^m }
		( - 1 )^{ ( \mathbf{s} \oplus \mathbf{a} \oplus \mathbf{b} \oplus \mathbf{c} ) \cdot \mathbf{x} }
		\ket{ \mathbf{a} }_{A} \ket{ \mathbf{b} }_{B} \ket{ \mathbf{c} }_{C}
		&=
		2^{m}
		\sum_{ \mathbf{b} \in \{ 0, 1 \}^m }
		\sum_{ \mathbf{c} \in \{ 0, 1 \}^m }
		\ket{ \mathbf{s} \oplus \mathbf{b} \oplus \mathbf{c} }_{A}
		\ket{ \mathbf{b} }_{B}
		\ket{ \mathbf{c} }_{C}
		\label{eq:Entangled Triple Sum Reduction - I}
		\\
		&=
		2^{m}
		\sum_{ \mathbf{a} \in \{ 0, 1 \}^m }
		\sum_{ \mathbf{c} \in \{ 0, 1 \}^m }
		\ket{ \mathbf{a} }_{A}
		\ket{ \mathbf{s} \oplus \mathbf{a} \oplus \mathbf{c} }_{B}
		\ket{ \mathbf{c} }_{C}
		\label{eq:Entangled Triple Sum Reduction - II}
		\\
		&=
		2^{m}
		\sum_{ \mathbf{a} \in \{ 0, 1 \}^m }
		\sum_{ \mathbf{b} \in \{ 0, 1 \}^m }
		\ket{ \mathbf{a} }_{A}
		\ket{ \mathbf{b} }_{B}
		\ket{ \mathbf{s} \oplus \mathbf{a} \oplus \mathbf{b} }_{C}
		\ .
		\label{eq:Entangled Triple Sum Reduction - III}
	\end{align}
	}
\end{tcolorbox}

By combining (\ref{eq:SEQSS$_3$ Phase 3 - I}) with (\ref{eq:Entangled Triple Sum Reduction - I}), (\ref{eq:Entangled Triple Sum Reduction - II}) and (\ref{eq:Entangled Triple Sum Reduction - III}), state $\ket{\psi_3}$ can be more succinctly as:

\begin{align} \label{eq:SEQSS$_3$ Phase 3 - II}
	\ket{\psi_3}
	&=
	\frac{1}{ 2^m }
	\sum_{ \mathbf{a} \in \{ 0, 1 \}^m }
	\sum_{ \mathbf{b} \in \{ 0, 1 \}^m }
	\sum_{ \mathbf{c} \in \{ 0, 1 \}^m }
	\ket{-}_{A}
	\ket{ \mathbf{a} }_{A}
	\ket{ \mathbf{b} }_{B}
	\ket{ \mathbf{c} }_{C}
	\ ,
\end{align}

where the states of the three players' Input Registers are always correlated as expressed by the following fundamental property

\begin{align} \label{eq:SEQSS$_3$ Fundamental Property}
	\mathbf{a} \oplus \mathbf{b} \oplus \mathbf{c} = \mathbf{s}
	\quad \Leftrightarrow \quad
	\mathbf{a} \oplus \mathbf{b} \oplus \mathbf{c} \oplus \mathbf{s} = \mathbf{0} \ .
\end{align}

Finally, Alice, Bob and Charlie measure their GHZ states in their Input Registers getting

\begin{align}
	\label{eq:SEQSS$_3$ Final Measurement}
	\ket{\psi_4}
	=
	\ket{ \mathbf{a} }_{A}
	\ket{ \mathbf{b} }_{B}
	\ket{ \mathbf{c} }_{C}
	\ , \quad \text{for some} \quad
	\mathbf{a}, \mathbf{b}, \mathbf{c} \in \{ 0, 1 \}^m
	\ ,
\end{align}

and thus, successfully completing the quantum part of the protocol.

We may easily observe that the contents of each of the three Input Registers appear random to Alice, Bob and Charlie, but are, when viewed as a composite system, correlated by the fundamental property (\ref{eq:SEQSS$_3$ Fundamental Property}) of the SEQSS$_3$ protocol. We now recall that Bod and Charlie are in the same region of space. This implies that they can securely exchange information without using any classical or quantum communication channel. A critical remark is that even if Bob and Charlie combine their measurements $\mathbf{b}$ and $\mathbf{c}$, they still will not be able to retrieve Alice's secret $\mathbf{s}$, unless, of course, it happens that $\mathbf{a} = \mathbf{0}$. This last event can happen with probability $\frac{1}{2^m}$, which is virtually negligible for large values of $m$. They still need a crucial ingredient from Alice, namely the contents $\mathbf{a}$ of Alice's Input Register. Hence, we can conclude the protocol by having Alice share her measurement $\mathbf{a}$ with \emph{anyone} of her agents via a public channel. She can choose either Bob or Charlie without affecting the protocol. Finally, Bob and Charlie, being in possecion of $\mathbf{a}$, they can combine their measurements and obtain the secret message $\mathbf{s}$, according to (\ref{eq:SEQSS$_3$ Fundamental Property}).

The part where Alice must share her measurement with her agents after the quantum part of the protocol, can be seen as an advantage, due to the fact that Alice and her agents may perform the quantum part of the protocol at a given time and then have Alice, as the spymaster, determine when will be the right time for her agents to unlock the secret message, by deciding when to broadcast her measurement.

Figure \ref{fig:Alice, Bob, Charlie and the Source of Entangled Triplets} provides a mnemonic visualization of the spatial positions of Alice, Bob and Charlie in the SEQSS$_3$ protocol, as well as the operation of the quantum and the classical channel.

\begin{tcolorbox}
	[
		grow to left by = 0.75 cm,
		grow to right by = 0.75 cm,
		colback = gray!01,
		enhanced jigsaw,		
		sharp corners,
		toprule = 1.0 pt,
		bottomrule = 1.0 pt,
		leftrule = 0.1 pt,
		rightrule = 0.1 pt,
		sharp corners,
		center title,
		fonttitle = \bfseries
	]
	\begin{figure}[H]
		\centering
		\begin{tikzpicture} [ scale = 1.0 ]
			\begin{scope}[on background layer]
				\node
				[
					rectangle, fill = Brown!25, minimum width = 75 mm, minimum height = 8 mm, anchor = center, label = { [ anchor = center ] center:Classical Public Channel $\mathbf{a}$ }
				]
				(Classical Channel) at ( 0.00, 3.50 ) {};
				\draw [Brown!25, ->, >=stealth, line width = 5.0 mm] (-3.50, 3.50) -- (-3.50, 1.50);
				\draw [Brown!25, ->, >=stealth, line width = 5.0 mm] (3.50, 3.50) -- (3.50, 1.50);
				\node
				[
					rectangle, fill = RedPurple!10, minimum width = 30 mm, minimum height = 30 mm, rounded corners = 10pt, anchor = center, label = { [ anchor = center ] center:\textbf{Alice} }
				]
				(Alice) at ( -3.50, 0.00 ) {};
				\node
				[
					rectangle, fill = WordDarkTealLighter80, minimum width = 14 mm, minimum height = 50 mm, anchor = center,
					label = { [ rotate = 90, align = center ] center:\textbf{Spatially}\\\textbf{Separated} }
				]
				(Space) at ( 0.00, 0.00 ) {};
				\node
				[
					circle, fill = GreenLighter2!10, minimum size = 30 mm, anchor = center, label = { [ anchor = center ] center:\textbf{Bob} }
				]
				(Bob) at ( 3.50, 0.00 ) {};
				\node
				[
					circle, fill = WordTurquoiseLighter80, minimum size = 30 mm, anchor = center, label = { [ anchor = center ] center:\textbf{Charlie} }
				]
				(Charlie) at ( 7.50, 0.00 ) {};
				\node
				[
					rectangle, fill = yellow!50, minimum width = 115 mm, minimum height = 10 mm, anchor = center,
					label = { [ align = center ] center:Quantum\\Channel }
				]
				(Quantum Channel) at ( 2.00, -3.52 ) {};
				\draw [ yellow!50, ->, >=stealth, line width = 5.0 mm ] (-3.50, -3.50) -- (-3.50, -1.50);
				\node [circle, fill = GreenLighter2, minimum size = 0.5 mm] () at (-3.50, -2.00) {  };
				\node () at (-2.75, -2.00) { $\ket{q_0}_{A}$  };
				\node () at (-3.50, -2.50) { $\vdots$ };
				\node [circle, fill = RedPurple, minimum size = 0.5 mm] () at (-3.50, -3.15) {  };
				\node () at (-2.50, -3.15) { $\ket{q_{m - 1}}_{A}$  };
				\draw [ yellow!50, ->, >=stealth, line width = 5.0 mm ] (3.50, -3.12) -- (3.50, -1.50);
				\node [circle, fill = GreenLighter2, minimum size = 0.5 mm] () at (3.50, -2.00) {  };
				\node () at (4.25, -2.00) { $\ket{q_0}_{B}$  };
				\node () at (3.50, -2.50) { $\vdots$ };
				\node [circle, fill = RedPurple, minimum size = 0.5 mm] () at (3.50, -3.15) {  };
				\node () at (4.50, -3.15) { $\ket{q_{m - 1}}_{B}$  };
				\draw [ yellow!50, ->, >=stealth, line width = 5.0 mm ] (7.50, -3.50) -- (7.50, -1.50);
				\node [circle, fill = GreenLighter2, minimum size = 0.5 mm] () at (7.50, -2.00) {  };
				\node () at (8.25, -2.00) { $\ket{q_0}_{C}$  };
				\node () at (7.50, -2.50) { $\vdots$ };
				\node [circle, fill = RedPurple, minimum size = 0.5 mm] () at (7.50, -3.15) {  };
				\node () at (8.50, -3.15) { $\ket{q_{m - 1}}_{C}$  };
				\node
				[
					rectangle, fill = WordAquaDarker50!90, text = white, minimum width = 20 mm, minimum height = 30 mm, align = center
				]
				(Source) at (2.00, -5.50) { \textbf{Source}\\\textbf{of} $GHZ_3$\\\textbf{triplets} };
			\end{scope}
		\end{tikzpicture}
		\caption{Alice is spatially separated from her agents Bob and Charlie, who are in the same region. A third party, the source, creates $m$ triplets of $GHZ_3$ entangled photons and sends one qubit from every triplet to Alice and the remaining two to Bob and Charlie.}
		\label{fig:Alice, Bob, Charlie and the Source of Entangled Triplets}
	\end{figure}
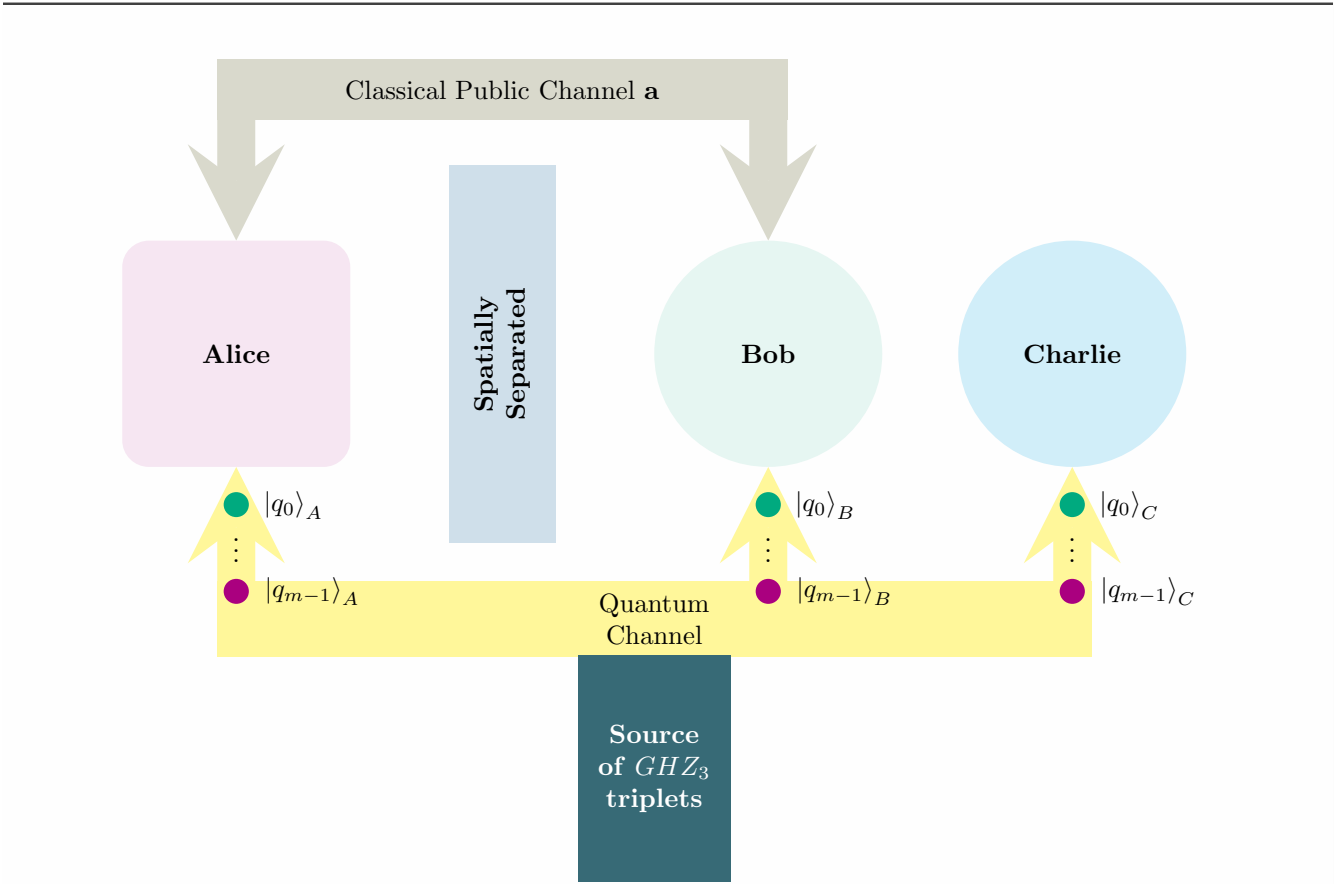
\end{tcolorbox}

\subsection{The $n$-player SEQSS$_n$ protocol} \label{sec:The n-Player SEQSS$_n$ Protocol}

Now that we have presented the simplest version of the protocol, we can move on to the general version, which can be described as an obvious extension of the SEQSS$_3$ game, where now Alice's agents are not only two but $n - 1$, namely Agent$_0$, \dots, Agent$_{ n - 2 }$. The quantum circuit implementing the SEQSS$_n$ protocol is depicted in the next Figure \ref{fig:The SEQSS$_n$ Protocol}. As before, Alice is spatially separated from her $n - 1$ agents, which are all located in the same region of space. Additionally, we point out that each of the $n$ players has her own quantum circuit and that the $n$ quantum circuits are similar, since they all contain an Input Register consisting of $m$ qubits. All the $n - 1$ Alice's secret agents have identical circuits. Alice's circuit differs because it also contains an Output Register consisting of a single qubit, upon which a Hadamard gate acts, and the unitary transform $U_{f_{A}}$ acting on both her Input and Output Registers. The corresponding qubits in the Input Registers used by Alice and her $n - 1$ secret agents constitute an $n$-tuple entangled in the $GHZ_n$ state. Table \ref{tbl:Figure SEQSS$_n$ Abbreviationss} explains the abbreviations that are used in the quantum circuit depicted in Figure \ref{fig:The SEQSS$_n$ Protocol}.

\begin{tcolorbox}
	[
		grow to left by = 0.75 cm,
		grow to right by = 0.75 cm,
		colback = gray!03,
		enhanced jigsaw,		
		sharp corners,
		toprule = 1.0 pt,
		bottomrule = 1.0 pt,
		leftrule = 0.1 pt,
		rightrule = 0.1 pt,
		sharp corners,
		center title,
		fonttitle = \bfseries
	]
	\centering
	\begin{figure}[H]
		\begin{center}
			\begin{tikzpicture} [ scale = 0.90 ]
				\begin{scope}[on background layer]
					\node
					[
						rectangle, fill = WordBlue!7, draw = black, dash pattern = on 0.30 cm off 0.25 cm, line width = 0.40 pt, minimum width = 145 mm, minimum height = 15 mm, anchor = west, label = { [ anchor = south ] south:\textbf{Agent}$_{ 0 }$ }
					]
					(agent n-1) at ( -1.5, -1.0 ) {};
					\node (dot 1) at ( 4.90, - 3.0 ) {\Large $\mathbf{\vdots}$};
					\node (dot 2) at ( 7.80, - 3.0 ) {\Large $\mathbf{\vdots}$};
					\node
					[
						rectangle, fill = GreenLighter2!10, draw = black, dash pattern = on 0.30 cm off 0.25 cm, line width = 0.40 pt, minimum width = 145 mm, minimum height = 15 mm, anchor = west, label = { [ anchor = south ] south:\textbf{Agent}$_{ n - 2 }$ }
					]
					(Bob) at ( -1.5, -5.25 ) {};
					\node
					[
						rectangle, fill = WordDarkTealLighter80, minimum width = 145 mm, minimum height = 10 mm, anchor = west, label = { [ anchor = center ] center:\textbf{Spatially Separated} }
					]
					(Space) at ( -1.5, -7.00 ) {};
					\node
					[
						rectangle, fill = RedPurple!10, draw = black, dash pattern = on 0.30 cm off 0.25 cm, line width = 0.40 pt, minimum width = 145 mm, minimum height = 30 mm, anchor = west, label = { [ anchor = south ] south:\textbf{Alice} }
					]
					(Alice) at ( -1.5, -9.50 ) {};
					\node
					[
						rectangle,  rotate = 90, rounded corners = 7 pt, fill = WordBlue, text = white, minimum width = 100 mm, minimum height = 10 mm
					]
					(Source) at (-3.00, -5.65) { \quad \textbf{Source of} $GHZ_n \ n$-\textbf{tuples} \quad };
					\draw [WordBlue, ->, >=stealth, line width = 3.0 mm] (-3.00, -1.05) -- (-1.25, -1.05);
					\draw [WordBlue, ->, >=stealth, line width = 3.0 mm] (-2.60, -5.25) -- (-1.25, -5.25);
					\draw [WordBlue, ->, >=stealth, line width = 3.0 mm] (-3.00, -8.70) -- (-1.25, -8.70);
				\end{scope}
				\begin{yquant}[ operator/separation = 0.24 cm, register/separation = 0.50 cm, every label/.append style = {align = center} ]
					%
					%
					nobit CAUX0;
					qubits { IR$_{ 0 }$ } CIR;
					nobit CAUX1;
					%
					%
					nobit BAUX0;
					nobit BAUX1;
					nobit BAUX2;
					qubits { IR$_{ n - 2 }$ } BIR;
					nobit BAUX3;
					%
					%
					nobit AAUX0;
					nobit AAUX1;
					qubits { AIR } AIR;
					qubit { $\ket{1}$ \\ AOR } AOR;
					nobit AAUX3;
					nobit AAUX4;
					[ name = Ph0, RedPurple, line width = 0.50 mm, label = Phase 0 ]
					barrier ( - ) ;
					h AOR; 
					[ name = Ph1, RedPurple, line width = 0.50 mm, label = Phase 1 ]
					barrier ( - ) ;
					%
					box {$ \ U_{f_{A}} \ $} (AIR, AOR);
					[ name = Ph2, RedPurple, line width = 0.50 mm, label = Phase 2 ]
					barrier ( - ) ;
					box { $H^{\otimes n}$ } CIR, BIR, AIR;
					[ name = Ph3, RedPurple, line width = 0.50 mm, label = Phase 3 ]
					barrier ( - ) ;
					measure CIR, BIR, AIR;
					[ name = Ph4, RedPurple, line width = 0.50 mm, label = Phase 4 ]
					barrier ( - ) ;
					hspace {0.3 cm} CIR;
					output {$\ket{ \mathbf{x}_{ n - 1 } }$} CIR;
					output {$\ket{ \mathbf{x}_{1} }$} BIR;
					output {$\ket{ \mathbf{a} }$} AIR;
					\node [ below = 6.00 cm ] at (Ph0) {$\ket{\psi_{0}}$};
					\node [ below = 6.00 cm ] at (Ph1) {$\ket{\psi_{1}}$};
					\node [ below = 6.00 cm ] at (Ph2) {$\ket{\psi_{2}}$};
					\node [ below = 6.00 cm ] at (Ph3) {$\ket{\psi_{3}}$};
					\node [ below = 6.00 cm ] at (Ph4) {$\ket{\psi_{4}}$};
				\end{yquant}
			\end{tikzpicture}
		\end{center}
		\caption{A schematic representation of the quantum circuit implementing the SEQSS$_n$ protocol.}
		\label{fig:The SEQSS$_n$ Protocol}
	\end{figure}
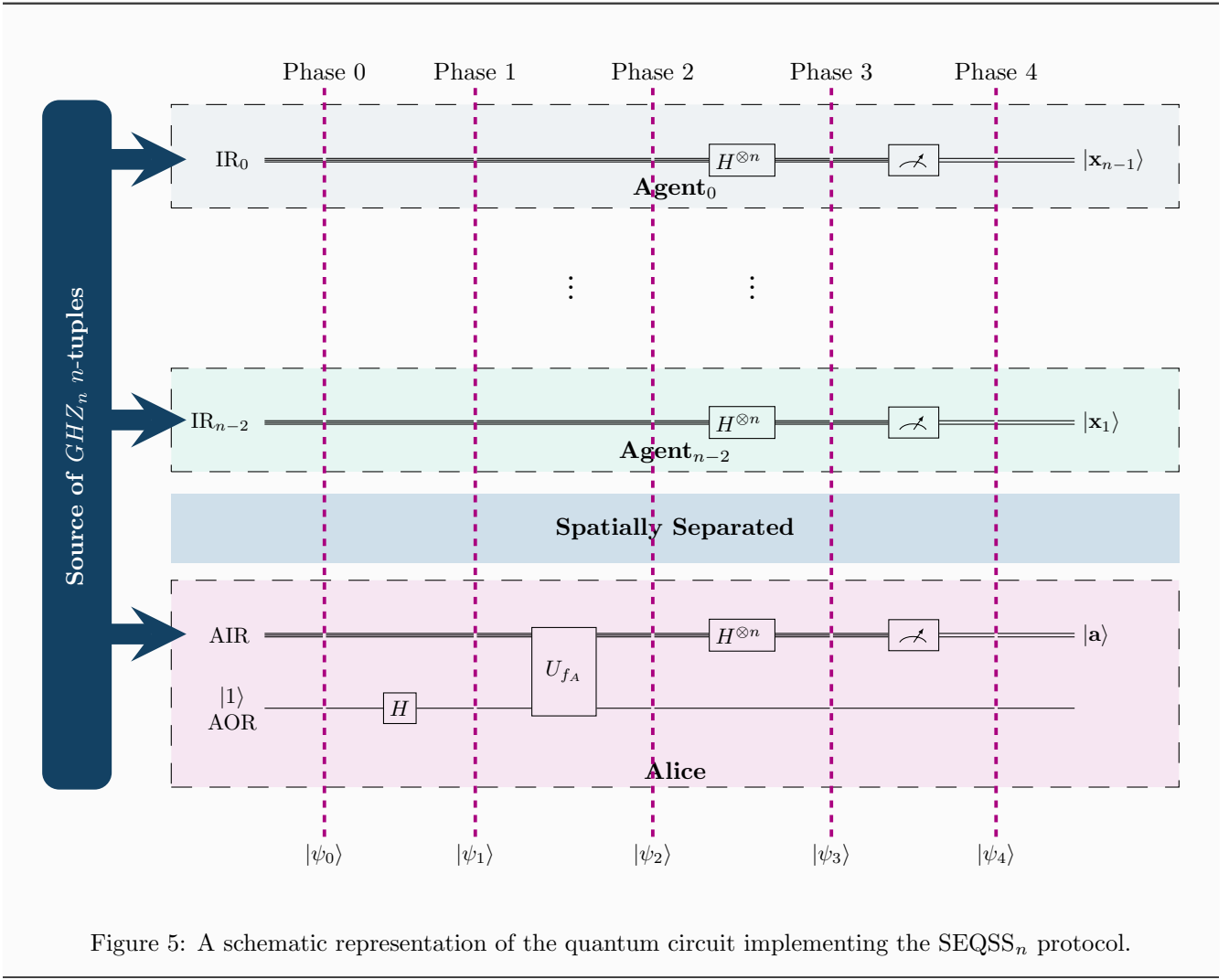
\end{tcolorbox}

\begin{table}[H]
	\renewcommand{\arraystretch}{1.40}
	\begin{tcolorbox}
		[
			grow to left by = -1.00 cm,
			grow to right by = -1.00 cm,
			colback = gray!03,
			enhanced jigsaw, 
			sharp corners,
			boxrule = 0.1 pt,
			toprule = 0.1 pt,
			bottomrule = 0.1 pt
		]
		\caption{This table shows the abbreviations that are used in Figure \ref{fig:The SEQSS$_n$ Protocol}. \\}
		\label{tbl:Figure SEQSS$_n$ Abbreviationss}
		\centering
		\begin{tabular}
			{
				>{\centering\arraybackslash} m{3.00 cm} !{\vrule width 1.25 pt}
				>{\centering\arraybackslash} m{5.00 cm} !{\vrule width 0.5 pt}
				>{\centering\arraybackslash} m{3.00 cm}
			}
			\Xhline{4\arrayrulewidth}
			\multicolumn{3}{c}{\textbf{Abbreviations} used in Figure \ref{fig:The SEQSS$_n$ Protocol}}
			\\
			\Xhline{\arrayrulewidth}
			\textbf{Abbreviations}
			&
			Full name
			&
			\# of qubits
			\\
			\Xhline{3\arrayrulewidth}
			AIR
			&
			Alice's Input Register
			&
			$m$
			\\
			\Xhline{1\arrayrulewidth}
			AOR
			&
			Alice's Output Register
			&
			$1$
			\\
			\Xhline{1\arrayrulewidth}
			IR$_{ i }, 0 \leq i \leq n - 2$
			&
			The Input Register of Agent$_i$
			&
			$m$
			\\
			\Xhline{4\arrayrulewidth}
		\end{tabular}
	\end{tcolorbox}
	\renewcommand{\arraystretch}{1.0}
\end{table}

Once again, following the steps of the circuit shown in Figure \ref{fig:The SEQSS$_n$ Protocol}, we can examine the steps of the algorithm more closely, by starting with the initial state of the system

\begin{align} \label{eq:SEQSS$_n$ Phase 0}
	\ket{ \psi_0 }
	=
	\frac{1}{ \sqrt{2^m} }
	\sum_{ \mathbf{x} \in \{ 0, 1 \}^m }
	\ket{1}_{A}
	\ket{\mathbf{x}}_{A} \ket{\mathbf{x}}_{ n - 2 } \dots \ket{\mathbf{x}}_{0}
	\ .
\end{align}

Similarly, $\ket{\mathbf{x}}_{A}$ gives the state of Alice's Input Register, $\ket{\mathbf{x}}_{i}, 0 \leq i \leq n - 2$, represent the states of the Input Registers of the $n - 1$ agents, and $\ket{1}_A$ is the state of Alice's Output Register. In our subsequent analysis the subscripts $A$, $0, 1, \dots, n - 2$ are consistently used to designate the registers belonging to Alice and Agent$_0$, \dots, Agent$_{ n - 2 }$, respectively. Thus, we may continue on to the next phase, by having Alice initiate the protocol, via the application of the Hadamard transform to her Output Register, which produces the ensuing state

\begin{align} \label{eq:SEQSS$_n$ Phase 1}
	\ket{\psi_1}
	=
	\frac{1}{ \sqrt{2^m} }
	\sum_{ \mathbf{x} \in \{ 0, 1 \}^m }
	\ket{-}_{A}
	\ket{\mathbf{x}}_{A} \ket{\mathbf{x}}_{ n - 2 } \dots \ket{\mathbf{x}}_{0}
	\ .
\end{align}

Akin to the previous version, this will allow Alice to apply her function given by (\ref{eq:Alice's Function}) on her registers and lead to the next state, which is

\begin{align} \label{eq:SEQSS$_n$ Phase 2 - II}
	\ket{\psi_2}
	&=
	\frac{ 1 }{ \sqrt{ 2^m } }
	\sum_{ \mathbf{x} \in \{ 0, 1 \}^m }
	( - 1 )^{ \mathbf{s} \cdot \mathbf{x} }
	\ket{-}_{A}
	\ket{\mathbf{x}}_{A} \ket{\mathbf{x}}_{ n - 2 } \dots \ket{\mathbf{x}}_{0}
	\ .
\end{align}

Afterwards, Alice and all her secret agents apply the $m$-fold Hadamard transformation to their Input Registers, driving the system into the next state

\begin{tcolorbox}
	[
		grow to left by = 1.50 cm,
		grow to right by = 1.00 cm,
		colback = white, 
		enhanced jigsaw, 
		sharp corners,
		boxrule = 0.01 pt,
		toprule = 0.01 pt,
		bottomrule = 0.01 pt
	]
	{\small
		\begin{align} \label{eq:SEQSS$_n$ Phase 3 - I}
			\ket{\psi_3}
			&=
			\frac{ 1 }{ \sqrt{ 2^m } }
			\sum_{ \mathbf{x} \in \{ 0, 1 \}^m }
			( - 1 )^{ \mathbf{s} \cdot \mathbf{x} }
			\ket{-}_{A}
			H^{ \otimes m } \ket{ \mathbf{x} }_{A}
			H^{ \otimes m } \ket{\mathbf{x}}_{ n - 2 } \dots H^{ \otimes m } \ket{\mathbf{x}}_{0}
			\nonumber \\
			&=
			\frac{ 1 }{ \sqrt{ 2^m } }
			\sum_{ \mathbf{x} \in \{ 0, 1 \}^m }
			( - 1 )^{ \mathbf{s} \cdot \mathbf{x} }
			\ket{-}_{A}
			\left( \frac{ 1 }{ \sqrt{ 2^m } } \sum_{ \mathbf{a} \in \{ 0, 1 \}^m } ( - 1 )^{ \mathbf{a} \cdot \mathbf{x} } \ket{ \mathbf{a} }_{A} \right)
			\nonumber \\
			&\left( \frac{ 1 }{ \sqrt{ 2^m } } \sum_{ \mathbf{y}_{ n - 2 } \in \{ 0, 1 \}^m } ( - 1 )^{ \mathbf{y}_{ n - 2 } \cdot \mathbf{x} } \ket{ \mathbf{y}_{ n - 2 } }_{ n - 2 } \right)
			\dots
			\left( \frac{ 1 }{ \sqrt{ 2^m } } \sum_{ \mathbf{y}_{ 0 } \in \{ 0, 1 \}^m } ( - 1 )^{ \mathbf{y}_{ 0 } \cdot \mathbf{x} } \ket{ \mathbf{y}_{ 0 } }_{0} \right)
			\nonumber \\
			&=
			\frac{ 1 }{ ( \sqrt{ 2^m } )^{ n + 1 } }
			\sum_{ \mathbf{x} \in \{ 0, 1 \}^m }
			\sum_{ \mathbf{a} \in \{ 0, 1 \}^m }
			\sum_{ \mathbf{y}_{ n - 2 } \in \{ 0, 1 \}^m }
			\dots
			\sum_{ \mathbf{y}_{ 0 } \in \{ 0, 1 \}^m }
			( - 1 )^{ ( \mathbf{s} \oplus \mathbf{a} \oplus \mathbf{y}_{ n - 2 } \oplus \cdots \oplus \mathbf{y}_{ 0 } ) \cdot \mathbf{x} }
			\ket{-}_{A}
			\ket{ \mathbf{a} }_{A} \ket{ \mathbf{y}_{ n - 2 } }_{ n - 2 } \dots \ket{ \mathbf{y}_{ 0 } }_{0}
			\ .
		\end{align}
	}
\end{tcolorbox}

The crucial observation now is that if

\begin{align}
	\mathbf{a} \oplus \mathbf{y}_{ n - 2 } \oplus \cdots \oplus \mathbf{y}_{ 0 } = \mathbf{s} \ ,
\end{align}

then $\forall  \mathbf{x} \in \{ 0, 1 \}^m$, the expression $( - 1 )^{ ( \mathbf{s} \oplus \mathbf{a} \oplus \mathbf{y}_{ n - 2 } \oplus \cdots \oplus \mathbf{y}_{ 0 } ) \cdot \mathbf{x} }$ becomes $( - 1 )^{0} = 1$. As a result, the sum $\sum_{ \mathbf{x} \in \{ 0, 1 \}^m } ( - 1 )^{ ( \mathbf{s} \oplus \mathbf{a} \oplus \mathbf{y}_{ n - 2 } \oplus \cdots \oplus \mathbf{y}_{ 0 } ) \cdot \mathbf{x} }$ $=$ $2^{m}$. Whenever $\mathbf{a} \oplus \mathbf{y}_{ n - 2 } \oplus \cdots \oplus \mathbf{y}_{ 0 } \neq \mathbf{s}$, the sum is just $0$ because for exactly half of the inputs $\mathbf{x}$ the exponent will be $0$ and for the remaining half the exponent will be $1$. Therefore, we derive that

\begin{align} \label{eq:General Inner Product Modulo 2 Property}
	\sum_{ \mathbf{x} \in \{ 0, 1 \}^m }^{}
	( - 1 )^{ ( \mathbf{s} \oplus \mathbf{a} \oplus \mathbf{y}_{ n - 2 } \oplus \cdots \oplus \mathbf{y}_{ 0 } ) \cdot \mathbf{x} }
	=
	2^{m} \delta_{ \mathbf{s}, \mathbf{a} \oplus \mathbf{y}_{ n - 2 } \oplus \cdots \oplus \mathbf{y}_{ 0 } }
	\ .
\end{align}

Equation (\ref{eq:General Inner Product Modulo 2 Property}) leads to the following $n$ equivalent and symmetric formulations.

\begin{tcolorbox}
	[
		grow to left by = 0.00 cm,
		grow to right by = 0.00 cm,
		colback = white, 
		enhanced jigsaw, 
		sharp corners,
		boxrule = 0.01 pt,
		toprule = 0.01 pt,
		bottomrule = 0.01 pt
	]
	{\small
		\begin{align}
			&\sum_{ \mathbf{a} \in \{ 0, 1 \}^m }
			\sum_{ \mathbf{y}_{ n - 2 } \in \{ 0, 1 \}^m }
			\dots
			\sum_{ \mathbf{y}_{ 0 } \in \{ 0, 1 \}^m }
			\sum_{ \mathbf{x} \in \{ 0, 1 \}^m }
			( - 1 )^{ ( \mathbf{s} \oplus \mathbf{a} \oplus \mathbf{y}_{ n - 2 } \oplus \cdots \oplus \mathbf{y}_{ 0 } ) \cdot \mathbf{x} }
			\ket{-}_{A}
			\ket{ \mathbf{a} }_{A} \ket{ \mathbf{y}_{ n - 2 } }_{ n - 2 } \dots \ket{ \mathbf{y}_{ 0 } }_{0}
			\nonumber \\
			&=
			2^{m}
			\sum_{ \mathbf{y}_{ n - 2 } \in \{ 0, 1 \}^m }
			\dots
			\sum_{ \mathbf{y}_{ 0 } \in \{ 0, 1 \}^m }
			\ket{-}_{A}
			\ket{ \mathbf{s} \oplus \mathbf{y}_{ n - 2 } \oplus \cdots \oplus \mathbf{y}_{ 0 } }_{A} \ket{ \mathbf{y}_{ n - 2 } }_{ n - 2 } \dots \ket{ \mathbf{y}_{ 0 } }_{0}
			\label{eq:Entangled n Sum Reduction - I}
			\\
			&=
			2^{m}
			\sum_{ \mathbf{a} \in \{ 0, 1 \}^m }
			\sum_{ \mathbf{y}_{ n - 3 } \in \{ 0, 1 \}^m }
			\dots
			\sum_{ \mathbf{y}_{ 0 } \in \{ 0, 1 \}^m }
			\ket{-}_{A}
			\ket{ \mathbf{a} }_{A} \ket{ \mathbf{s} \oplus \mathbf{a} \oplus \mathbf{y}_{ n - 3 } \oplus \cdots \oplus \mathbf{y}_{ 0 } }_{ n - 2 } \ket{ \mathbf{y}_{ n - 3 } }_{ n - 3 } \dots \ket{ \mathbf{y}_{ 0 } }_{0}
			\label{eq:Entangled n Sum Reduction - II}
			\\
			& \hspace{7 cm} \dots
			\nonumber \\
			&=
			2^{m}
			\sum_{ \mathbf{a} \in \{ 0, 1 \}^m }
			\sum_{ \mathbf{y}_{ n - 2 } \in \{ 0, 1 \}^m }
			\dots
			\sum_{ \mathbf{y}_{ 1 } \in \{ 0, 1 \}^m }
			\ket{-}_{A}
			\ket{ \mathbf{a} }_{A}
			\ket{ \mathbf{y}_{ n - 2 } }_{ n - 2 } \dots \ket{ \mathbf{y}_{ 1 } }_{1}
			\ket{ \mathbf{s} \oplus \mathbf{a} \oplus \mathbf{y}_{ n - 2 } \oplus \cdots \oplus \mathbf{y}_{ 1 } }_{ 0 }
			\label{eq:Entangled n Sum Reduction - III}
		\end{align}
	}
\end{tcolorbox}

By combining (\ref{eq:SEQSS$_n$ Phase 3 - I}) with (\ref{eq:Entangled n Sum Reduction - I}), (\ref{eq:Entangled n Sum Reduction - II}) and (\ref{eq:Entangled n Sum Reduction - III}), we can write state $\ket{\psi_3}$ in a compact way as

\begin{align} \label{eq:SEQSS$_n$ Phase 3 - II}
	\ket{\psi_3}
	&=
	\frac{ 1 }{ ( \sqrt{ 2^m } )^{ n - 1 } }
	\sum_{ \mathbf{a} \in \{ 0, 1 \}^m }
	\sum_{ \mathbf{y}_{ n - 2 } \in \{ 0, 1 \}^m }
	\dots
	\sum_{ \mathbf{y}_{ 0 } \in \{ 0, 1 \}^m }
	\ket{-}_{A}
	\ket{ \mathbf{a} }_{A} \ket{ \mathbf{y}_{ n - 2 } }_{ n - 2 } \dots \ket{ \mathbf{y}_{ 0 } }_{0}
	\ ,
\end{align}

where the states of all the $n$ players' Input Registers are always correlated as dictated by the fundamental property

\begin{align} \label{eq:SEQSS$_n$ Fundamental Property}
	\mathbf{a} \oplus \mathbf{y}_{ n - 2 } \oplus \cdots \oplus \mathbf{y}_{ 0 } = \mathbf{s}
	\quad \Leftrightarrow \quad
	\mathbf{a} \oplus \mathbf{y}_{ n - 2 } \oplus \cdots \oplus \mathbf{y}_{ 0 } \oplus \mathbf{s} = \mathbf{0} \ .
\end{align}

Finally, Alice and her secret agents Agent$_0$, \dots, Agent$_{ n - 2 }$ measure their GHZ states in their Input Registers getting

\begin{align}
	\label{eq:SEQSS$_n$ Final Measurement}
	\ket{\psi_4}
	=
	\ket{ \mathbf{a} }_{A}
	\ket{ \mathbf{y}_{ n - 2 } }_{ n - 2 } \dots \ket{ \mathbf{y}_{ 0 } }_{0}
	\ , \quad \text{for some} \quad
	\mathbf{a}, \mathbf{y}_{ 0 }, \dots, \mathbf{y}_{ n - 2 } \in \{ 0, 1 \}^m
	\ ,
\end{align}

which completes the quantum part of the SEQSS$_n$ protocol.

In this most general case, we encounter the same motif we saw in the case of the three players. In particular, although the contents of each of the $n$ Input Registers appear random to Alice and her secret agents, when viewed as a composite system, they are correlated by the fundamental property (\ref{eq:SEQSS$_n$ Fundamental Property}) of the SEQSS$_n$ protocol. The fact that all of Alice's agents are in the same geographical region implies that they can securely exchange information without using any classical or quantum communication channel. However, even if all the $n - 1$ secret agents combine their measurements $\mathbf{y}_{ 0 }, \dots, \mathbf{y}_{ n - 2 }$, they will still not be able to retrieve Alice's secret $\mathbf{s}$, unless, of course, it happens that $\mathbf{a} = \mathbf{0}$. This last event can happen with probability $\frac{1}{2^m}$, which tends to zero as $m$ increases. They lack a crucial ingredient from Alice, namely the contents $\mathbf{a}$ of Alice's Input Register. Hence, we can conclude the protocol by having Alice share her measurement $\mathbf{a}$ with \emph{anyone} of her $n - 1$ agents via a public channel. She can choose either one of them without affecting the protocol. Finally, the $n - 1$ secret agents, being in possecion of $\mathbf{a}$, they can combine their measurements and obtain the secret message $\mathbf{s}$, according to (\ref{eq:SEQSS$_n$ Fundamental Property}).

Let us again stress that Alice may share her measurement with \emph{anyone} of her agents. This is undoubtedly an important advantage of the protocol, due to the fact that Alice and her agents may perform the quantum part of the protocol at a given time and then have Alice as the spymaster, determine when will be the right time for her agents to unlock the secret message, by deciding when to broadcast her measurement. Furthermore, the broadcast of Alice’s measurement via a public channel, will not hinder the security of the protocol because even if Eve is present, she will still need the rest of the measurements, in order to retrieve the secret message $\mathbf{s}$.

Figure \ref{fig:Alice, her Secret Agents and the Source of Entangled n-tuples} presents a mnemonic representation of the SEQSS$_n$ protocol, together with the operation of the quantum and the classical channel.

\begin{tcolorbox}
	[
		grow to left by = 0.75 cm,
		grow to right by = 0.75 cm,
		colback = gray!01,
		enhanced jigsaw,		
		sharp corners,
		toprule = 1.0 pt,
		bottomrule = 1.0 pt,
		leftrule = 0.1 pt,
		rightrule = 0.1 pt,
		sharp corners,
		center title,
		fonttitle = \bfseries
	]
	\begin{figure}[H]
		\centering
		\begin{tikzpicture} [ scale = 1.0 ]
			\begin{scope}[on background layer]
				\node
				[
					rectangle, fill = Brown!25, minimum width = 75 mm, minimum height = 8 mm, anchor = center, label = { [ anchor = center ] center:Classical Public Channel $\mathbf{a}$ }
				]
				(Classical Channel) at ( 0.00, 3.50 ) {};
				\draw [Brown!25, ->, >=stealth, line width = 5.0 mm] (-3.50, 3.50) -- (-3.50, 1.50);
				\draw [Brown!25, ->, >=stealth, line width = 5.0 mm] (3.50, 3.50) -- (3.50, 1.50);
				\node
				[
					rectangle, fill = RedPurple!10, minimum width = 30 mm, minimum height = 30 mm, rounded corners = 10pt, anchor = center, label = { [ anchor = center ] center:\textbf{Alice} }
				]
				(Alice) at ( -3.50, 0.00 ) {};
				\node
				[
					rectangle, fill = WordDarkTealLighter80, minimum width = 14 mm, minimum height = 50 mm, anchor = center,
				label = { [ rotate = 90, align = center ] center:\textbf{Spatially}\\\textbf{Separated} }
				]
				(Space) at ( 0.00, 0.00 ) {};
				\node
				[
					circle, fill = GreenLighter2!10, minimum size = 30 mm, anchor = center, label = { [ anchor = center ] center:\textbf{Agent}$_{ n - 2 }$ }
				]
				(Bob) at ( 3.50, 0.00 ) {};
				\node (Dots) at ( 6.00, 0.00 ) {\Large \textbf{\dots}};
				\node
				[
					circle, fill = WordTurquoiseLighter80, minimum size = 30 mm, anchor = center, label = { [ anchor = center ] center:\textbf{Agent}$_{ 0 }$ }
				]
				(Charlie) at ( 8.50, 0.00 ) {};
				\node
				[
					rectangle, fill = yellow!50, minimum width = 125 mm, minimum height = 10 mm, anchor = center,
					label = { [ align = center ] center:Quantum\\Channel }
				]
				(Quantum Channel) at ( 2.50, -3.52 ) {};
				\draw [ yellow!50, ->, >=stealth, line width = 5.0 mm ] (-3.50, -3.50) -- (-3.50, -1.50);
				\node [circle, fill = GreenLighter2, minimum size = 0.5 mm] () at (-3.50, -2.00) {  };
				\node () at (-2.75, -2.00) { $\ket{q_0}_{A}$  };
				\node () at (-3.50, -2.50) { $\vdots$ };
				\node [circle, fill = RedPurple, minimum size = 0.5 mm] () at (-3.50, -3.15) {  };
				\node () at (-2.50, -3.15) { $\ket{q_{m - 1}}_{A}$  };
				\draw [ yellow!50, ->, >=stealth, line width = 5.0 mm ] (3.50, -3.12) -- (3.50, -1.50);
				\node [circle, fill = GreenLighter2, minimum size = 0.5 mm] () at (3.50, -2.00) {  };
				\node () at (4.25, -2.00) { $\ket{q_0}_{n - 2}$  };
				\node () at (3.50, -2.50) { $\vdots$ };
				\node [circle, fill = RedPurple, minimum size = 0.5 mm] () at (3.50, -3.15) {  };
				\node () at (4.50, -3.15) { $\ket{q_{m - 1}}_{n - 2}$  };
				\draw [ yellow!50, ->, >=stealth, line width = 5.0 mm ] (8.50, -3.50) -- (8.50, -1.50);
				\node [circle, fill = GreenLighter2, minimum size = 0.5 mm] () at (8.50, -2.00) {  };
				\node () at (9.25, -2.00) { $\ket{q_0}_{0}$  };
				\node () at (8.50, -2.50) { $\vdots$ };
				\node [circle, fill = RedPurple, minimum size = 0.5 mm] () at (8.50, -3.15) {  };
				\node () at (9.50, -3.15) { $\ket{q_{m - 1}}_{0}$  };
				\node
				[
					rectangle, fill = WordAquaDarker50!90, text = white, minimum width = 20 mm, minimum height = 30 mm, align = center
				]
				(Source) at (2.50, -5.50) { \textbf{Source}\\\textbf{of} $GHZ_n$\\$n$-\textbf{tuples} };
			\end{scope}
		\end{tikzpicture}
		\caption{Alice is spatially separated from her $n - 1$ agents, who are in the same region. A possibly different entity, the source, creates $m$ $n$-tuples of $GHZ_n$ entangled photons and sends one qubit from every $n$-tuple to Alice and her $n - 1$ secret agents.}
		\label{fig:Alice, her Secret Agents and the Source of Entangled n-tuples}
	\end{figure}
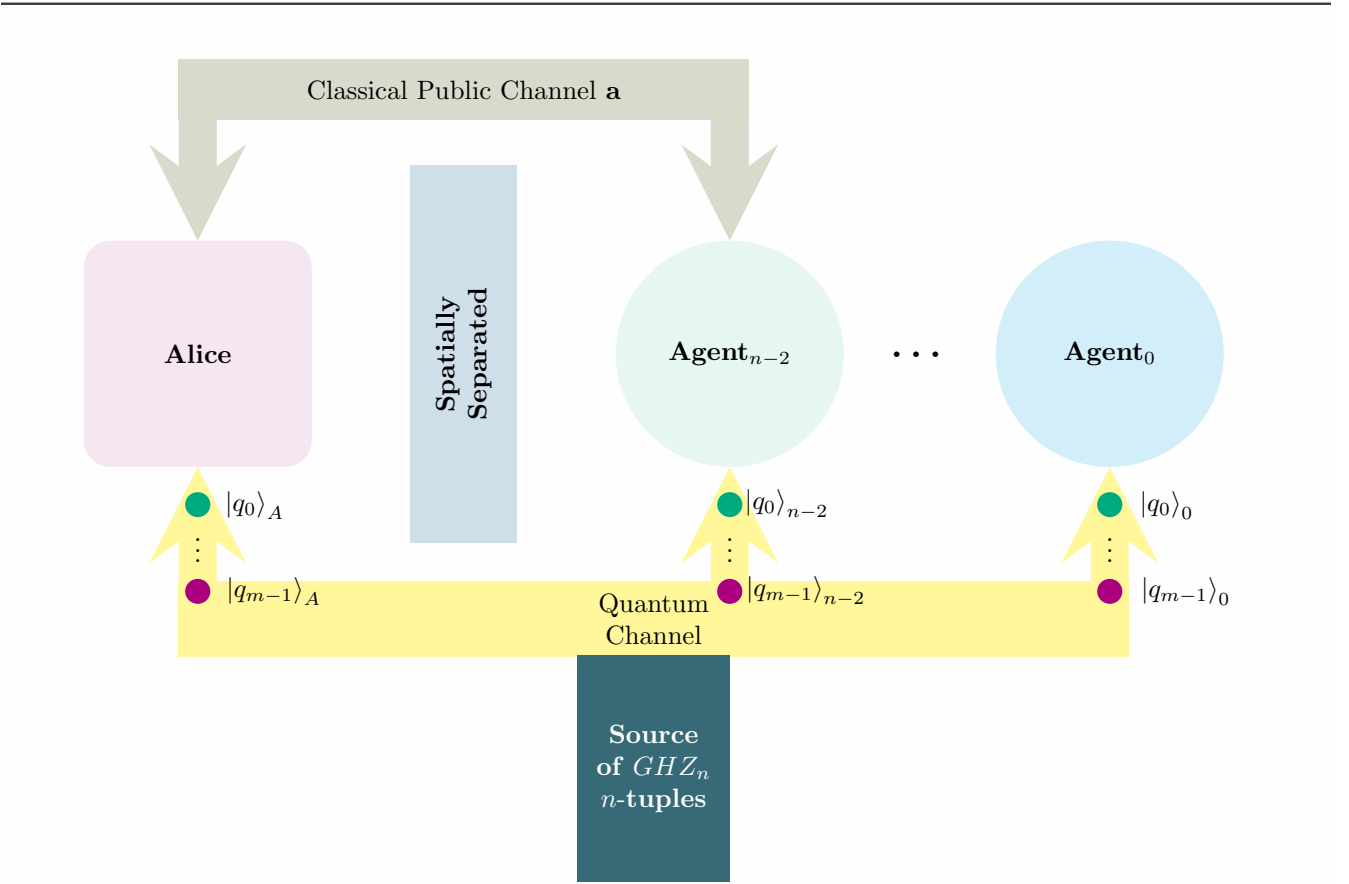
\end{tcolorbox}
\section{Discussion and conclusions} \label{sec:Discussion and Conclusions}

In this paper, we proposed a new entanglement based QSS protocol, called SEQSS$_n$, that relies on the use of maximally entangled GHZ tuples, evenly distributed among the players, giving Alice the spymaster the ability to securely share a secret message with her agents, in a simple and symmetric way, since all her agents are treated similarly, utilizing identical quantum circuits. We presented in great detail the simplest version of the protocol and afterwards we continued our analysis with the general version with $n$ players, thus proving its scalability to as many players as we want. Moreover, we showed that our proposal can give a rather useful advantage to Alice that plays the role of the spymaster and is responsible for the transmission of the secret message $\textbf{s}$, by giving her the ability to decide when the rest of the players will have access to the message, even at a time instant after the completion of the quantum part of the protocol. We will close this discussion with a brief example of how Alice's added advantage can be used and with an elementary security analysis of the protocol.

A relatively interesting proposition on the effective utilization of our spymaster's added advantage can be described as a clandestine mission with Alice as the spymaster and the rest of her agents acting on her behalf. In this mission, Alice has created a detailed plan, outlining any possible scenario of the mission and she has divided it, into multiple secret messages, i.e., $\textbf{s}_1, \textbf{s}_2, \dots$. At the beginning of the mission, Alice and her agents complete the quantum part of the protocol multiple times, in order for everyone to have a piece of every single secret message. At a later moment, Alice can decide on when or if she wants to broadcast some of her measurements based on the real time info of her agents. For example she will broadcast her piece of information $\textbf{s}_1$ to start the mission, and then, based on how the mission progresses, she will only broadcast in real time only her pieces of the secret messages that are relevant to that scenario of the mission. Thus, providing her with the ability to hide from her agents and the enemy possible critical information. Furthermore, since she can broadcast her pieces of the secret message, after the completion of the quantum part, on a public channel, Alice and her agents can use whatever communication method they deem suitable for the mission.

Finally, let us emphasize that the security of our protocol is attributed to its entanglement based nature, as the property of Entanglement Monogamy does not allow the entanglement of a maximally entangled tuple with any other qubit. This in turn prevents Eve from gaining any information by trying to entangle a qubit of the GHZ tuples used in our protocol, during the transmission of the GHZ tuples to the players. Additionally, Alice's broadcast of her piece of the secret message to her agents, will not hinder the security of protocol, because Eve must somehow gather all the pieces from all the agents in order to be able to acquire the secret message. However, we believe that as a future work, a more exhaustive and detailed security analysis of the protocol, along with its performance analysis, will prove beneficial to further test and guarantee its reliability.

\bibliographystyle{ieeetr}
\bibliography{QuantumSecretSharing}

\end{document}